\def\a{\alpha}\def\b{\beta}\def\d{\delta}\def\e{\epsilon}
\def\f{\phi}\def\g{\gamma}\def\h{\theta}
\def\k{\kappa}\def\l{\lambda}\def\m{\mu}\def\n{\nu}\def\r{\rho}\def\s{\sigma}
\def\y{\eta}
\def\ee{\varepsilon}

\def\D{\Delta}

\def\mo{{-1}}\def\ha{{1\over 2}}

\def\({\left(}\def\){\right)}\def\[{\left[}\def\]{\right]}
\def\lra{\leftrightarrow}\def\bdot{\!\cdot\!}

\def\mn{{\mu\nu}}

\def\tran{transformations }\def\coo{coordinates }

\def\rep{representation }

\def\poi{Poincar\'e }

\def\wrt{with respect to }\def\ie{i.e.\ }

\def\cor{commutation relations }

\def\section#1{\bigskip\noindent{\bf#1}\smallskip}

\def\PL#1{Phys.\ Lett.\ {\bf#1}}
\def\PRL#1{Phys.\ Rev.\ Lett.\ {\bf#1}}
\def\PR#1{Phys.\ Rev.\ {\bf#1}}\def\CQG#1{Class.\ Quantum Grav.\ {\bf#1}}
\def\NP#1{Nucl.\ Phys.\ {\bf#1}}

 \def\IJMP#1{Int.\ J. Mod.\ Phys.\ {\bf #1}}
\def\MPL#1{Mod.\ Phys.\ Lett.\ {\bf #1}}

\def\JHEP#1{JHEP\ {\bf#1}}

\def\LRR#1{Liv.\ Rev.\ Rel.\ {\bf#1}}
\def\hep#1{{\tt hep-th/#1}}

\def\ref#1{\medskip\everypar={\hangindent 2\parindent}#1}
\def\beginref{\begingroup
\bigskip
\centerline{\bf References}
\nobreak\noindent}
\def\endref{\par\endgroup}

\def\bdot{\!\cdot\!}
\newfam\msbfam
\font\tenmsb=msbm10  \textfont\msbfam=\tenmsb
\def\Bbb{\fam\msbfam \tenmsb}
\def\cA{{\cal A}}\def\cC{{\cal C}}\def\cH{{\cal H}}\def\cU{{\cal U}}\def\cM{{\cal M}}
\def\ot{\otimes}\def\op{\oplus}\def\idt{{\rm id}}\def\cx{{\Bbb C}}

\magnification=1200

{\nopagenumbers
\line{}
\vskip30pt
\centerline{\bf Generalized relativistic kinematics in Poincar\'e-invariant models}

\vskip60pt
\centerline{
{\bf B. Iveti\'c}$^{1,}$\footnote{$^\dagger$}{e-mail: boris.ivetic@irb.hr}
{\bf S. Mignemi}$^{2,3,}$\footnote{$^\ddagger$}{e-mail: smignemi@unica.it},
and {\bf A. Samsarov}$^{1,}$\footnote{$^*$}{e-mail: samsarov@unica.it},}
\vskip10pt
\smallskip
\centerline{$^1$Rudjer Bo\v skovi\'c Institute, Bijeni\v cka cesta 54, 10002 Zagreb, Croatia}
\smallskip
\centerline{$^2$Dipartimento di Matematica e Informatica, Universit\`a di Cagliari}
\centerline{viale Merello 92, 09123 Cagliari, Italy}
\centerline{$^3$INFN, Sezione di Cagliari, Cittadella Universitaria, 09042 Monserrato, Italy}
\vskip80pt
\centerline{\bf Abstract}
\medskip
{Assuming the validity of the relativity principle, we discuss the implications on relativistic kinematics of a
deformation of the \poi invariance that preserves the \poi algebra, and only modifies its action on spacetime
in a Lorentz-invariant way.
We show that, in contrast to the case where the \poi algebra is deformed, the action of boosts on two-particle
states is not affected, while the addition law of momenta is to a large extent arbitrary.
We give some nontrivial examples of this arising from doubly special relativity and noncommutative geometry
and show that Hopf-algebra methods give equivalent results.}
\vskip10pt
{\noindent

\vfil\eject}}

\section{1. Introduction}
Most theories of quantum gravity predict the existence of a new fundamental scale at which the effects of gravitation
and quantum theory should merge, which is usually identified with the Planck scale ($L\approx 10^{-33}$cm, or equivalently
$M\approx 10^{-11}$g). In particular, the existence
of a minimal measurable length is a natural consequence of the introduction of such scale [1].

The new scale may affect special relativity by deforming the \poi invariance in such a way that the observer independence
of the laws of physics is not spoiled.
This is the idea at the origin of doubly special relativity (DSR) [2]. The formalism of DSR is based on the analysis
of momentum space, and in particular of the deformation of the standard energy-momentum dispersion law, and essentially
relies on relativistic classical mechanics.
This approach has lead to several interesting developments, as the hypothesis that momentum space has a constant curvature
[3], and that as a consequence locality may depend on the observer [4].

A different approach to the fundamental scale is based instead on the analysis of position space, and has lead to the
investigation of noncommutative geometries (NCG) [5], that are founded on the assumption that spacetime has a quantum
nature at the Planck scale, and hence positions cannot be sharply measured. Among the different approaches to NCG, the
most interesting in the context of relativistic kinematics are those based on the formalism of Hopf algebras [6], since
they describe the effects of noncommutativity on the spacetime symmetries.
Several models have been studied in this formalism, among which
the Moyal plane [7], the Snyder geometry [8-10] and especially the $\k$-Minkowski spacetime with its associated
$\k$-\poi algebra [11-13].

All these theories are strictly connected and have in common the deformation of the Heisenberg algebra of commutation
relations between position and momentum variables, hence leading to generalized uncertainty relations [14].
Usually they are also associated with a nonlinear deformation of the action of the Lorentz algebra on momentum space,
(and hence a deformation of the \poi symmetry). Examples are given by many DSR models [2,15] or, for what concerns NCG,
by $\k$-\poi models [11-12].

It has been shown that the deformation of the \poi invariance has nontrivial effects on the relativistic kinematics.
In fact, the deformed transformations of momenta may not agree with a linear addition law and hence it may be
necessary to deform also the addition law of momenta. However, there is no unique way to define the new composition
law starting from the deformation of the Lorentz symmetry, and some additional assumption must be made, that can
lead to rather varied physical predictions. Some proposals are based on classical notions, with the introduction of
auxiliary momentum variables that transform in the standard way [16] and give rise to simple rules for the composition.
For models based on the Hopf algebra formalism, instead, the computations are highly nontrivial, and the addition law
may be noncommutative and nonassociative [12,17].

More involved kinematical effects may also arise: for example, in models exhibiting a noncommutative
addition law of momenta, the action of Lorentz boosts on the momentum
of a particle in a two-particle system can be deformed in such a way to depend on the momenta of both particles.
This fact was first noticed in refs.\ [18,19] for DSR models inspired by the $\k$-\poi formalism.

With the aim of clarifying these issues,
in ref.\ [20] the proposal was advanced to derive the most general deformation of relativistic kinematics compatible
with a given DSR model and in accordance with a relativity principle, \ie with the observer independence of the
laws of physics. The analysis was based on elementary postulates and a classical formalism, without reference to a
specific framework.
It was performed for generic Lorentz-invariant models with deformed boost action on momentum space, but
preserving the rotational invariance, and was carried out up to order $1/M$ in the deformation scale, identified
with the inverse of Planck mass.
It turned out that some relations occur between the deformation parameters of the dispersion relation, of the
addition of momenta law, and of the boosts, so that the deformations depend altogether on five free parameters.

In more detail, the analysis of ref.\ [20] relies on the following assumptions: the rotational invariance is respected;
the deformed dispersion relation must be invariant under the deformed boosts; the composition law of momenta may be
nonlinear, and in particular, in the case of a two-particle system the transformation of each particle can depend on
the momentum of the other;
finally, the transformation under boosts of the total momentum of a two-particle system must be equal to the
addition law of the boosted momenta of the single particles, so that the energy-momentum
conservation holds for any inertial observer (relativity principle).

In this context, however, an important case was overlooked: in fact, it is known that  DSR and NCG do not necessarily
imply the deformation of the \poi algebra.
Actually, there are models that preserve the full \poi algebra, deforming only its action on spacetime coordinates
in an explicitly Lorentz-invariant fashion.
Among these the best known is the one proposed by Snyder [8] in the 1940s, and more recently generalized in ref.\ [9,10].
This model is based on a deformation of the Heisenberg algebra that leaves the \poi algebra invariant, introducing
the noncommutativity of the spacetime coordinates. Several aspects of this model have been discussed in the literature
in recent years [10,17,21,22].

The purpose of this paper is to extend the investigations of [20] to this class of models, where the \poi invariance
is not deformed, or is deformed in a minimal (\ie Lorentz-invariant\footnote{$^1$}{By Lorentz-invariant we mean invariant
under linear Lorentz transformations.}) way. This assumption still leaves room for a
deformation of relativistic kinematics analogous to the one occurring in the case of deformed \poi algebra. The
deformation can again be expanded in powers of the scale $1/M$. However, in this case the leading corrections
can only arise to order $1/M^2$, since we require that they preserve the (undeformed) Lorentz invariance, and therefore
were not considered in [20].
We expect significant differences from the case studied there of deformations of the \poi symmetry that only preserve
rotational invariance, since the constraints are stronger. In fact, we show that in our case the two-particle \tran
are not modified, but still there is a great freedom in the choice of the addition law of momenta.

The paper is organized as follows:
in sect.\ 2 we discuss the relativity principle for systems with undeformed \poi symmetry;
in sect.\ 3 we extend the discussion to the case where the \poi symmetry is deformed by Lorentz-invariant terms;
in sect.\ 4 we show that our results are consistent with kinematical constraints on photon decay;
in sect.\ 5 we discuss some explicit examples of nontrivial addition laws arising from a classical approach, while
in sect.\ 6 we discuss our problem from the point of view of Hopf algebras, obtaining equivalent results.

\section{2. Relativity principle}
Our study is based on the same hypotheses of ref.\ [20], listed above, to which is added the request that the
\poi algebra is unaltered and all deformations preserve Lorentz invariance.
We restrict our study to systems with no more than two particles, and to leading order in $1/M$, and,
taking the point of view of DSR, discuss the effects of the symmetries on momentum space.

We start our investigation by considering a deformation of special relativity that preserves the action of boosts
on one-particle states, but possibly deforms the addition law of momenta and the action of boosts
on two-particle states in a Lorentz-invariant way.
We assume that the deformations can be expanded in powers of $1/M$ and consider only leading order corrections.

Hence, the one-particle infinitesimal boost \tran $L(p)$ of parameters $\e_i$ on the momenta $p_\m$ coincide with those of
special relativity:\footnote{$^2$}{In the following we use greek indices for spacetime, latin indices for spatial variables
and denote $p\bdot q=p^\m q_\m$ and $p^2=p^\m p_\m$.}
$$\big[L(p)\big]_0=p_0+\e_ip_i,\qquad\big[L(p)\big]_i=p_i+\e_ip_0.\eqno(1)$$
The other kinematical relations may instead be deformed by Lorentz-invariant terms. This request implies that the leading order
corrections are $o(1/M^2)$.

The general form of the dispersion law of a particle of mass $m$ compatible with Lorentz invariance will of course be
a function of the Casimir invariant $p^2$,
$$C[p]=p^2+{\a\over M^2}\,p^4+o\({1\over M^4}\)=m^2,\eqno(2)$$
with $\a$ a free parameter. Alternatively, eq.\ (2) can be written in the standard form $p^2=\m^2$, where $\m$ is an effective
mass defined as $\m^2=m^2(1-\a m^2/M^2)$.

We consider now the scattering of two particles. The composition law of their momenta $p_\m$ and $q_\m$ that satisfies
our assumptions, to leading order has the general form
$$p_\m\oplus q_\m=p_\m+q_\m+{1\over M^2}(\b_1p^2+\b_2p\bdot q+\b_3 q^2)p_\m+{1\over M^2}(\g_1q^2+\g_2p\bdot q+\g_3 p^2)q_\m,
\eqno(3)$$
with free parameters $\b_i$, $\g_i$.

As noticed in [18,19], the boost \tran of a component of a two-particle system may include terms depending on the momentum of
the other particle. This results from specific examples  in DSR or $\k$-\poi models.
The general form of the \tran compatible with Lorentz invariance will be
$$L^+_q(p)=L(p)+\bar L^+_q(p),\qquad L^-_p(q)=L(q)+\bar L^-_p(q),\eqno(4)$$
with
$$\eqalign{&\big[\bar L^+_q(p)\big]_0={\e_i\over M^2}\[(\y_2^+ p\bdot q+\y_3^+ q^2)p_i+(\h_1^+ q^2+\h_2^+ p\bdot q+\h_3^+ p^2)q_i\],\cr
&\big[\bar L^+_q(p)\big]_i={\e_i\over M^2}\[(\r_2^+ p\bdot q+\r_3^+ q^2)p_0+(\s_1^+ q^2+\s_2^+ p\bdot q+\s_3^+ p^2)q_0\],\cr
&\big[\bar L^-_p(q)\big]_0={\e_i\over M^2}\[(\y_2^- p\bdot q+\y_3^- p^2)q_i+(\h_1^- p^2+\h_2^- p\bdot q+\h_3^- q^2)p_i\],\cr
&\big[\bar L^-_p(q)\big]_i={\e_i\over M^2}\[(\r_2^- p\bdot q+\r_3^- p^2)q_0+(\s_1^- p^2+\s_2^- p\bdot q+\s_3^- q^2)p_0\].}\eqno(5)$$
The $\y_i^\pm$, $\h_i^\pm$, $\r_i^\pm$ and $\s_i^\pm$ are free parameters and the $\pm$ superscripts refer to the fact that the two
particles may have different transformation properties.
The terms $p^2$ and $q^2$ are of course the effective masses of the two particles, cf.\ the comment after eq.\ (2).

The invariance of the dispersion relation of each of the particles under boosts requires that
$C[L^+_q(p)]=C[p]$, $C[L^-_p(q)]=C[q]$, and then $C[\bar L_q^+(p)]=C[\bar L_p^-(q)]=0$. This implies that
$$\b_1=\g_1=0\eqno(6)$$
and
$$\y_2^\pm=\r_2^\pm, \qquad \y_3^\pm=\r_3^\pm,\qquad \h_k^\pm=\s_k^\pm=0\quad (k=1,2,3).\eqno(7)$$
Actually, condition (6) is equivalent to the natural requirement that
$$p\oplus q|_{q=0}=p,\qquad p\oplus q|_{p=0}=q.\eqno(8)$$
Therefore, the boost transformation of a two-particle system depends on four independent parameters $\y_2^\pm$,
$\y_3^\pm$. These conditions are independent of the value of $\b_i$ and $\g_i$.

Finally, the relativity principle requires that
$$L(p\oplus q)=L_q^+(p)\oplus L_p^-(q).\eqno(9)$$
This guarantees the invariance under boosts of the energy-momentum conservation in the decay of one particle into two.
A short calculation shows that (9) implies that $\y_2^\pm=\r_2^\pm=\y_3^\pm=\r_3^\pm=0$.
It follows that under our assumptions the standard form of the Lorentz invariance
is preserved also at the two-particle level.

Hence, as one may have expected, the request that the  Lorentz \tran of one-particle states are not deformed implies
that also those of two-particle states are unchanged.
On the other hand, the composition law of momenta may be deformed and depends in general on four parameters
$\b_2$, $\b_3$, $\g_2$ and $\g_3$, namely,
$$p_\m\oplus q_\m=p_\m+q_\m+{1\over M^2}(\b_2p\bdot q+\b_3 q^2)p_\m+{1\over M^2}(\g_2p\bdot q+\g_3 p^2)q_\m.\eqno(10)$$
Also the dispersion law is not constrained, but can be any function of $p^2$.

These results are in contrast with the case when the deformation of the \poi invariance does not preserve explicitly
the Lorentz symmetry [20].
In that case the composition law of momenta depends on five arbitrary parameters, while the boost \tran of one-
and two-particle states, as well as the dispersion law, depend in a definite way from these and two further parameters.

It is easy to extend to all orders in $1/M$ our proof that if one-particle \tran are undeformed, also the action of
boosts on two-particle states is not deformed, and that no restrictions occur on the form of the momenta composition law,
except Lorentz invariance.

\section{3. Minimal Lorentz deformation}
We consider now a more general case, where the action of the Lorentz \tran on the momentum of a single particle is
deformed, but only by Lorentz-invariant terms.

To leading order, the one-particle boosts are now
$$\big[L(p)\big]_0=p_0+\e_i\(1+{\y_1\over M^2}p^2\)p_i,\qquad\big[L(p)\big]_i=p_i+\e_i\(1+{\y_1\over M^2}p^2\)p_0\eqno(11)$$
with $\y_1$ a new parameter. The deformation is therefore proportional to the effective mass $\m^2$ of the particle.
The deformation of the two-particle boosts instead is still of the general form (5).

The requirement of the invariance of the dispersion relation of each of the particles under boosts again enforces
conditions (6)-(7), while (9) now implies $\y_1=\ha\y_2^\pm=\y_3^\pm$.
Therefore, the boosts for the two-particle system read in this case
$$\eqalign{&L^+_q(p)_0=p_0+\e_i\[1+{\y_1\over M^2}(p+q)^2\]p_i,\qquad L^+_q(p)_i=p_i+\e_i\[1+{\y_1\over M^2}(p+q)^2\]p_0,\cr
&L^+_p(q)_0=q_0+\e_i\[1+{\y_1\over M^2}(p+q)^2\]q_i,\qquad L^+_p(q)_i=q_i+\e_i\[1+{\y_1\over M^2}(p+q)^2\]q_0,}\eqno(12)$$
and depend therefore on the unique parameter $\y_1$ that deforms the one-particle sector, and are symmetric in $p$ and $q$.
Hence, as in the previous case, two-particle boosts are completely determined by one-particle transformations.
Analogously, the composition law of the momenta is still independent of the deformation of the boosts and depends on four
parameters.

Although the generalization considered in this section can be of some interest, in the following of the paper we shall
concentrate on the more common case of undeformed \poi invariance investigated in sect.\ 2.

\section{4. Kinematical constraints}
The consistency of the addition law (10) can be tested by verifying that it does not violate simple
kinematical constraints on particle decays valid in special relativity.
Following ref.\ [19], we may for example verify that it is compatible with the request that photon decay is not allowed.

Let us consider the process $\g\to e^+\,e^-$, and denote $p_\g$, $p_+$ and $p_-$, the 4-momenta of
the photon, the positron and the electron respectively, with $p=(E,{\bf p})$.
In special relativity, one has
$$p_\g=p_++p_-.\eqno(13)$$
Since the mass of the photon vanishes, the square of this relation gives
$$(p_++p_-)^2=2(m^2+p_+\bdot p_-)=0,\eqno(14)$$
with $m$ the mass of the electron. But
$$p_+\bdot p_-=E_+E_--|{\bf p}_+|\,|{\bf p}_-|\cos\h\sim E_+E_-(1-\cos\h)-{m^2\over2}\({E_+\over E_-}+
{E_-\over E_+}\)\cos\h,\eqno(15)$$
where $\h$ is the angle between the outgoing particles, and we have used the ultrarelativistic approximation for
the momenta of the massive particles.
Then,
$$\cos\h\sim{E_+E_-+m^2\over E_+E_--{m^2\over2}\({E_+\over E_-}+{E_-\over E_+}\)},\eqno(16)$$
which is always greater than 1, so that the process is not allowed.

In the case of deformed addition of momenta (10), the relation (14) becomes
$$0=(p_+\oplus p_-)^2=2(\m^2+p_+\bdot p_-)\[1+{\b_2+\g_2\over M^2}\,p_+\bdot p_-+2{\b_3+\g_3\over M^2}\,\m^2\],\eqno(17)$$
where $\m$ is the effective mass defined in sect.\ 2.
From (17) one obtains with straightforward calculations,\footnote{$^3$}{Actually, eq.\ (17) admits also
the solution $p_+\bdot p_-=-(M^2+2(\b_3+\g_3)\m^2)/(\b_2+\g_2)$, that can however be excluded on physical grounds,
since it does not have the correct limit for vanishing deformation parameters.}
$$p_+\bdot p_-=-\m^2\eqno(18)$$
exactly as in the classical case, except for the presence of the effective mass.
Hence, the relation (16) is still valid and the process of photon decay is not allowed kinematically.

Analogous results can be obtained for more complicated processes, as for example pair production in photon-photon
scattering and so on.

\section{5. Snyder model and its generalizations}
We give now some explicit examples of nontrivial theories in which the \poi algebra is preserved, and show that the
addition law of momenta always takes the form (10), with suitable parameters.

The best known nontrivial example of Poincar\'e-invariant model is given by the Snyder model [8].
In this model, the action of the Lorentz algebra, with generators $J_\mn=x_\m p_\n-x_\n p_\m$, on
positions and momenta is undeformed,\footnote{$^4$}{The classical version of the model is simply obtained by replacing
the commutators with Poisson brackets.}
$$\eqalign{&\ [J_\mn,J_{\r\s}]=i\(\y_{\m\r}J_{\n\s}-\y_{\m\s}J_{\n\r}-(\mu\lra\nu)\),\qquad[p_\m,p_\n]=0,\cr
&[J_\mn,p_\l]=i\(\y_{\m\l}p_\n-\y_{\l\n}p_\m\),\qquad[J_\mn,x_\l]=i\(\y_{\m\l}x_\n-\y_{\n\l}x_\m\),}\eqno(19)$$
while the action of the translations, generated by $p_\m$, on positions is deformed and implies, due to the Jacobi
identities, the noncommutativity of the spacetime \coo $x_\m$:
$$[x_\m,p_\n]=i\(\y_\mn+{p_\m p_\n\over M^2}\),\qquad[x_\m,x_\n]=i{J_\mn\over M^2}.\eqno(20)$$

The model can be generalized [9], preserving the \poi invariance (19) and modifying (20) in a Lorentz-invariant way as
$$[x_\m,p_\n]=i\(f(p^2/M^2)\y_\mn +g(p^2/M^2)\,{p_\m p_\n\over M^2}\),\qquad[x_\m,x_\n]=i{J_\mn\over M^2},
\eqno(21)$$
where $f(p^2/M^2)$ and $g(p^2/M^2)$ are functions of $p^2/M^2$, that, in order to obey the Jacobi identities,
must satisfy the relation
$$g={1+2ff'\over f-2{p^2\over M^2}f'},\eqno(22)$$
where the prime denotes a derivative \wrt $p^2/M^2$. The Snyder model is recovered for $f(p^2/M^2)=1$.

The \cor (21) can be realized in terms of commutative coordinates $X_\m$, obeying canonical \cor with the $p_\m$, by
setting [9]
$$x_\m=X_\m f+{1\over M^2}\ X\bdot p\,p_\m g,\eqno(23)$$
where a specific operator ordering has been chosen, and the Lorentz generators are now given by $J_\mn=X_\m p_\n-X_\n p_\m$.

A further slight generalization of the model can be obtained by defining new momenta $P_\m=h(p^2/M^2)p_\m$. This only
changes the
form of the \cor $[x_\m,P_\n]$ and the explicit realization of the $J_\mn$, but leaves the algebra otherwise unaltered.
However, if one allows this possibility, several different representations can be obtained in terms of canonical coordinates
for given \cor (19), (21). For example, the original \rep of the Snyder \cor was given, in terms of canonical
\coo $X_\m$, $P_\m$, by (cf.\ (23))
$$p_\m=P_\m,\qquad x_\m=X_\m+{X\bdot P\over M^2}\,P_\m,\eqno(24)$$
but a different \rep has been introduced in [22],
$$p_\m={P_\m\over\sqrt{1-P^2/M^2}},\qquad x_\m=\sqrt{1-P^2/M^2}\ X_\m.\eqno(25)$$
\medskip

At this point, it is important to remark that the
deformed Heisenberg and \poi algebras are not sufficient to describe the physics of a DSR model, but it is also necessary
to specify an addition law for the momenta compatible with the deformation. The law is not uniquely defined for a given
deformation, and several proposals have been advanced for giving a prescription. We are not interested here in discussing
the physical plausibility of the different proposals, but we
just require their compatibility with the relativity principle stated above.

We first discuss a framework, introduced in ref.\ [16], that relies on purely classical reasoning and gives by construction
a commutative addition law. It is based on the fact that, as discussed above,
the phase space \coo $x_\m$ and $p_\m$ can be related by a Darboux transformation to canonical variables $X_\m$ and $P_\m$,
which in this framework are interpreted just as auxiliary variables.
In particular, one can choose \tran of the form $p_\m=F_\m(P_\m)$, and assume that the auxiliary momenta $P_\m$, $Q_\m$ add
linearly, $P_\m\oplus Q_\m=P_\m+Q_\m$. One then goes back to the physical variables $p_\m$, $q_\m$ to get the nonlinear
addition rule
$$p_\m\oplus q_\m=F\big(F^\mo(p_\m)+F^\mo(q_\m)\big).\eqno(26)$$
Since the Darboux transformation is not unique, the result will depend on the specific choice of representation,
\ie of the $P$'s and $Q$'s.

For the original Snyder transformation (24), the definition (26) of course gives the classical addition rule,
$p_\m\oplus q_\m=p_\m+q_\m$.

For the transformation (25), one obtains instead to leading order,
$$p_\m\oplus q_\m\sim\(1+{q^2\over2M^2}+{q\bdot p\over M^2}\) p_\m+\(1+{p^2\over2M^2}+{q\bdot p\over M^2}\)q_\m\eqno(27)$$
This is an example of our general formula (10) with $\b_2=\g_2=1$, $\b_3=\g_3=\ha$.

It is easy to see that for general $F$ the prescription (26) always gives $\b_2=\g_2=2\b_3=2\g_3$. A short calculation
shows that these are also the conditions that ensure that the addition law (26) is associative to leading order in $1/M$.
It must be noticed, however, that the axioms of sect.\ 2 do not necessarily imply that the three-particle addition law can
be deduced from the one holding for two particles.

\section{6. Generalized Snyder model and Hopf algebra}
A different approach to the addition rule of momenta for generalized Snyder models is that of noncommutative geometry.
This is  based on the formalism of Hopf algebras, of which we recall some basic definitions [6].

Let us consider an algebra $\cA$ on $\cx$ with elements $a_i$, and a product $\cA\ot\cA\to\cA$ denoted by
$m(a_1\ot a_2)=a_1a_2$. The algebra is associative,
$$m(a_1\ot m(a_2\ot a_3))=m(m(a_1\ot a_2)\ot a_3),\eqno(28)$$
and has unit $I$. We denote $\y$ the linear map from $\cx$ to $\cA$ such that $\y(1)=I$.

A coalgebra $\cC$ is a vector space where a coproduct $\D:\cC\to\cC\ot\cC$ and a counit
$\ee:\cC\to\cx$ are defined, which satisfy the axioms:
$$\eqalignno{&(\D\ot\idt)\circ\D=(\idt\ot\D)\circ\D,\qquad\qquad{\rm (coassociativity)}&(29)\cr
&(\ee\ot\idt)\circ\D=(\idt\ot\ee)\circ\D=\idt,&(30)}$$
where $\idt$ is the identity map on $\cC$.

A bialgebra $\cH$ is an algebra which is also a coalgebra. The compatibility of the two structures
requires that
$$\eqalignno{&\D(a_1a_2)=\D(a_1)\D(a_2),&(31)\cr
&\ee(a_1a_2)=\ee(a_1)\ee(a_2).&(32)}$$

Finally, a bialgebra equipped with a linear map  $S:\cH\to\cH$, called antipode, which is antihomomorphic,
$S(a_1a_2)=S(a_2)S(a_1)$ and satisfies
$$m(S\ot\idt)\circ\D=m(\idt\ot S)\circ\D=\y\circ\ee,\eqno(33)$$
is called a Hopf algebra.

Hopf algebras can be used to describe the action of a group of
transformations on multiparticle states.
More precisely, the algebraic sector describes the symmetries of the
physical space in which particles  move, while the coalgebraic
sector gives the rules for the action of the symmetry group on a tensor product
of particle states.

To complete  the mathematical setting we must introduce a module algebra $\cM$ which includes all
physical states and in general all functions of the commutative coordinates $X$.
The action of the symmetry algebra $\cH$ on the
module $\cM$ is then realized through the action\
$\triangleright: \cH \otimes \cM \rightarrow
\cal{M} $, as $(a,\f) \mapsto a \triangleright\f$, for any
element $a$ in $\cH$ and any commutative function $\f(X)$
in $\cal{M}$. In our case the algebras of interest are the \poi
algebra $iso(1,3)$ (cf.\ (19)) and the corresponding universal enveloping algebra
$\cA\equiv\cU(iso(1,3))$.
We can promote $\cA$ to a Hopf algebra by introducing the counits
$\varepsilon (p_\mu) = \varepsilon (J_\mn) = 0$ and the
coproducts for the generators $p_\m$ and $J_\mn$.\footnote{$^5$}{The
  coproduct for the translation generators is given by eq.\ (37), while
  the coproduct for Lorentz generators can be taken to be a primitive
  one (although its form is not essential for the subsequent analysis).}

The action of $\cA$ on $\cal{M}$ is given by the standard prescription,
$p_\mu \triangleright \phi = -i\partial_\mu \phi$, i.e. $J_{\mu \nu }
\triangleright \phi = -i (X_\mu \partial_\nu - X_\nu \partial_\mu ) \phi.$
The compatibility between the algebra structure in  $\cal{M}$ and
the  $\cH$ action is provided by the requirement
$$ a \triangleright \bigg( m_*(\phi_1 \otimes \phi_2  ) \bigg) =  m_* \bigg[ \Delta (a)
 \triangleright (\phi_1 \otimes \phi_2  ) \bigg],\eqno(34)$$
where $m_*$ is the so called star product, a noncommutative
multiplication in the module algebra  $\cM$ induced by the
noncommutative nature of Snyder space. The element $a$ can be any element in
$\cA$ and $\phi_1, \phi_2$ any
two elements in $\cM$. The axiom (34) enforces the Leibniz rule.
\medskip
In this paper we are in particular interested in the action of the translation group on particle states in
noncommutative Snyder spacetime.
The coproduct describes then the way in which the group elements act on two-particle states and hence how
the momenta add, and depends on the star product through the compatibility axioms stated above, while
 the antipode plays the role of the inverse transformation for the group action.

The Hopf algebra associated to the generalized Snyder models can be obtained starting from the realization (23).
We do not report here the details of the calculation, but refer to [10]. In particular, to get the addition law of
momenta, we need the coproduct.
This can be calculated to leading order by expanding the function $f$ in (23) in powers of $1/M$, namely,
$f=1+c\,p^2/M^2+o(1/M^4)$, with $c$ an arbitrary parameter. To leading order one gets
$$\eqalign{\D p_\m=p_\mu\ot I+I\ot p_\m&+{1+4c\over2M^2}\,p_\m p_\n\ot p^\n+{c\over M^2}\,p_\m\ot p^2\cr
&+{1+2c\over M^2}\,p_\n\ot p^\n  p_\m+{1+2c\over2M^2}\,p^2\ot p_\m.}\eqno(35)$$

From (35) the addition law of momenta follows at once. It has of course the form (10), with
$$\b_2=\ha+2c,\quad\b_3=c,\quad\g_2=1+2c,\quad\g_3=\ha+c.\eqno(36)$$
This is in accordance with our result that, even if the \poi algebra is not deformed, the addition law of momenta
can be modified. For this class of models, the deformation of the algebra depends to leading order on a single
parameter $c$.
In particular, for the Snyder model in the original realization (24), $c=0$, while in the realization (25) $c=-\ha$.

It may be interesting to investigate if more general models based on Hopf algebras can admit addition laws depending
on a larger number of parameters, to match the general law (10). With this aim, we investigate the constraints that
Hopf algebra axioms impose on the general form of the Lorentz-invariant addition law of momenta.

Starting from the ansatz (3), we can write down the corresponding coproduct,
$$\eqalign{\D p_\mu  = p_\mu \otimes I + I \otimes p_\mu  + &
  {\beta_1\over M^2} p^2 p_\mu \otimes I + {\beta_2\over M^2} p_\mu
  p_\nu \otimes p^{\nu} + {\beta_3\over M^2} p_\mu \otimes  p^2 \cr
+ & {\gamma_1\over M^2} I \otimes p^2 p_\mu   +  {\gamma_2\over M^2} p_\nu
   \otimes p^\nu  p_{\mu} + {\gamma_3\over M^2}    p^2 \otimes p_\mu.}\eqno(37)$$
From this it is possible to extract the associated realization of the generalized Snyder model, or equivalently,
the commutation relations $[x_\m, p_\n]$.
These can be found by imposing the compatibility condition (34), which yields
$$ [x_\m, p_\n] =i\y_\mn\( 1+ {\b_3\over M^2}p^2\) - {\g_1\over M^2}x_\m p^2 p_\n+
i{\g_2\over M^2} p_\m  p_\n +O \({1\over M^4}\).\eqno(38)$$
If one moreover requires $[p_\m,p_\n]=0$, the \cor (38) are compatible with the Jacobi identities only if $\g_1=0$.
In that case, the Jacobi identities also imply
$$[x_\m,x_\n]=i{\g_2-2\b_3\over M^2}\,J_\mn,\eqno(39)$$
which is compatible with (21) only if
$$\g_2-2\b_3=1.\eqno(40)$$

One can now enforce axiom (30), obtaining the conditions
$$\b_1=0,\qquad\g_1=0,\eqno(41)$$
that coincide with the constraints (6) of sect.\ 2.
Note that the second condition has also been obtained from (38) assuming commuting momentum components.

If we further require that the coassociativity condition (29) is satisfied, we get three additional nontrivial
conditions
$$\b_2=\g_2=2\b_3=2\g_3.\eqno(42)$$
The same constraints for an associative addition law were also obtained by elementary methods at the end of sect.\ 5.
It is easily seen that the conditions (42) are not compatible with (36).
This shows that the coproduct in Snyder space cannot be coassociative.
However, this fact is not surprising since coassociativity implies the associativity of the star product,
and the fact that the star product in Snyder space is not associative has been already known for some time [10,17].

If we do not require coassociativity and commutativity of momenta,
we are left with four free parameters $\b_2$, $\b_3$, $\g_2$ and $\g_3$.
These are the same obtained from general considerations on the implementation of the relativity principle in sect.\ 2.
It appears therefore that the Hopf algebra treatment gives the same results as the general formalism based on the
relativity principle, and that the model defined by the \cor (19) and (21) is not the most general one compatible with
the relativity principle, since to leading order it leads to a one-parameter addition law, eq.\ (36).
In fact, the relations (21) can be further generalized preserving Lorentz invariance and momentum commutativity, giving
rise to a two-parameter addition law at linearized level [23].

A further extension of the model could possibly be obtained by
not imposing the condition of commutativity of the momenta in (19), since this is not required for the validity of the
general expression (10). However, changing this
assumption would include position-dependent terms in the commutators and this prevents the possibility of using
the formalism of ref.\ [10] to calculate the addition law of the momenta. This problem is currently under study.

The number of free parameters can still be reduced by imposing some additional requirements.
For example, a natural request is that the antipode for $p_\mu$ be undeformed (at least to order $1/M^2$).
Application of the axiom (33) then translates into the requirement
$$ p_\m\op S(p_\m) = S(p_\m) \op p_\m = 0.\eqno(43)$$
When this is applied to  eq.\ (3), one obtains the condition
$$\eqalign{p_\m+S(p_\m)&+
 {1\over M^2}\bigg( \b_1 p^2  +  \b_2 p \bdot S(p) + \b_3 {S(p)}^2 \bigg) p_\m\cr
& +{1\over M^2}\bigg(\g_1 {S(p)}^2+\g_2 p \bdot S(p) +  \g_3 p^2\bigg) S(p_\m) =0,}\eqno(44)$$
and analogously for the the second relation. The ansatz
$$  S(p_\m) = -p_\mu+{\d\over M^2} p^2 p_\m +O ({1\over M^4})\eqno(45)$$
leads to  $ \d = \b_2 + \g_1 - \b_1 -\b_3 + \g_3 - \g_2$.
Therefore, the requirement that the antipode is undeformed yields
$$\d=0=\b_2-\b_3+\g_3-\g_2.\eqno(46)$$
It is evident that eq.\ (43) is equivalent to the natural requirement that, in the formalism of sect.\ 2,
$p_\m\op(-p_\m)=0$,
\ie that the total momentum of two particles with opposite momenta vanishes. Of course, this request also leads
to the condition (46).

\section{7. Conclusion}
In this paper we have investigated the deformations of the addition law and of the boost transformations of momenta
compatible with the principle of relativity in Poincar\'e-invariant models of deformed relativity.
Previous investigations concerned models with deformed \poi invariance, which enjoy rather different properties [20].

Our results can be summarized as follows:
if the action of the Lorentz group on the momentum of a one-particle state is preserved, also that on two-particle
states is. This is true to any order in perturbation theory.
If instead the action on one-particle states is deformed by Lorentz-invariant terms, also that on two-particle states
is deformed in a definite way.

Moreover, in both cases the addition law of the momenta can be deformed and to leading order can depend on four free
parameters.
These results are confirmed by the fact that a deformed addition law of this kind does not spoil the kinematical
constraints on photon decay of special relativity.

We have given explicit examples of deformed addition laws, coming from DSR and NCG, based on the Snyder model and
its generalizations, and have shown that also the formalism of the Hopf algebra leads to the same constraints
obtained from the relativity principle.
The possibility that the addition law of momenta can be deformed at the Planck scale without violating
the relativity principle and the \poi invariance deserves further investigations.

\section{Acknowledgements}
\noindent We wish to thank Jos\'e Carmona for several useful comments and discussions.
SM wishes to thank Stjepan Meljanac and Rina \v Strajn for enlightening  discussions.
BI would like to thank the Universit\`a di Cagliari for kind hospitality.

\noindent The research leading to these results  has received funding from the European Union Seventh Framework Programme
(FP7 2007-2013) under grant agreement n° 291823 Marie Curie FP7-PEOPLE-2011-COFUND – NEWFELPRO. This work was
partially supported by the European Commission and the Croatian Ministry of Science, Education and Sports
 as a part of a project "Noncommutative geometry, gravity and black holes (NCGGBH)" which has received funding
 through NEWFELPRO project under grant agreement n° 63.

\bigskip

\beginref
\ref [1] L.J. Garay, \IJMP{A10}, 145 (1995); S. Hossenfelder, \LRR{16}, 2 (2013).
\ref [2] G. Amelino-Camelia, \PL{B510}, 255 (2001), \IJMP{D11}, 35 (2002).
\ref [3] J. Kowalski-Glikman, \PL{B547}, 291 (2002); J. Kowalski-Glikman, S. Nowak, \CQG{20}, 4799 (2003).
\ref [4] G. Amelino-Camelia, L. Freidel, J. Kowalski-Glikman and L. Smolin, \PR{D84}, 084010 (2011).
\ref [5] S. Doplicher, K. Fredenhagen and J.E. Roberts, \PL{B331}, 39 (1994).
\ref [6] S. Majid, {\it Foundation of quantum group theory}, Cambridge University Press 1995;
A.P. Balachandran, S.G. Jo and G. Marmo, {\it Group theory and Hopf algebras}, World Scientific 2010.
\ref [7] V.P. Nair and A.P. Polychronakos, \PL{B505}, 267 (2001);
 L. Mezinescu, \hep{0007046}.
\ref [8] H.S. Snyder, \PR{71}, 38 (1947).
\ref [9] M.V. Battisti and S. Meljanac, \PR{D79}, 067505 (2009).
\ref [10] S. Meljanac, D. Meljanac, A. Samsarov and M. Stojic, \MPL{A25}, 579 (2010); \PR{D83}, 065009 (2011);
 M.V. Battisti and S. Meljanac, \PR{D82}, 024028 (2010);
\ref [11] J. Lukierski, H. Ruegg, A. Novicki and V.N. Tolstoi, \PL{B264}, 331 (1991);
J. Lukierski, A. Novicki and H. Ruegg, \PL{B293}, 344 (1992).
\ref [12] J. Kowalski-Glikman and Nowak, \PL{B539}, 126 (2002);
J. Lukierski and A. Nowicki, \IJMP{A} 18, 7 (2003).
\ref [13] G. Amelino-Camelia and M. Arzano, \PR{D65}, 084044 (2002);
A. Agostini, G. Amelino-Camelia, F. d'Andrea, \IJMP{A19}, 5187 (2004).
\ref [14] M. Maggiore, \PL{B304}, 63 (1993); \PL{B319}, 83 (1993).
\ref [15] J. Magueijo and L. Smolin, \PRL{88}, 190403 (2002).
\ref [16] S. Judes and M. Visser, \PR{D68}, 045001 (2003).
\ref [17] F. Girelli and E.L. Livine, \JHEP{1103}, 132 (2011); AIP Conf.\ Proc.\  {\bf 1196}, 115 (2009).
\ref [18] G. Gubitosi and F. Mercati, \CQG{20}, 145002 (2013).
\ref [19] G. Amelino-Camelia, \PR{D85}, 084034 (2012).
\ref [20] J.M. Carmona, J.L. Cort\'es and F. Mercati, \PR{D86}, 084032 (2012).
\ref [21] R. Banerjee, S. Kulkarni and S. Samanta, \JHEP{0605}, 077 (2006);
J.M. Romero and A. Zamora, \PL{B661}, 11 (2008);
H.Y Guo, C.G. Huang, Y. Tian, H.T. Wu, Z Xu and B. Zhou, \CQG{24}, 4009 (2007);
L. Lu and A. Stern, \NP{B854}, 894 (2011).
\ref [22] S. Mignemi, \PR{D84}, 025021 (2011).
\ref [23] S. Meljanac, private communication.
\endref

\end